\documentclass[prl,reprint,superscriptaddress]{revtex4-1}

\usepackage{graphicx}
\usepackage{bm}
\usepackage{hyperref}
\usepackage{color}
\usepackage{easyReview}
\usepackage{latexsym} 
\usepackage{ulem}
\begin{document}
	
	\title{Interaction between Surface Acoustic Wave and Quantum Hall Effects}
	
	\date{today}
	
\author{Xiao Liu}
\affiliation{International Center for Quantum
	Materials, Peking University, Beijing, China, 100871}
\author{Mengmeng Wu}
 \affiliation{International Center for Quantum
	Materials, Peking University, Beijing, China, 100871}
\author{Renfei Wang} 
\affiliation{International Center for Quantum
	Materials, Peking University, Beijing, China, 100871}
\author{Xinghao Wang}
 \affiliation{International Center for Quantum
	Materials, Peking University, Beijing, China, 100871}
\author{Wenfeng Zhang} 
\affiliation{International Center for Quantum
	Materials, Peking University, Beijing, China, 100871}
\author{Yujiang Dong} 
\affiliation{International Center for Quantum
	Materials, Peking University, Beijing, China, 100871}

\author{Rui-Rui Du}
\affiliation{International Center for Quantum
	Materials, Peking University, Beijing, China, 100871}
\affiliation{Center for Excellence, University of Chinese Academy of Sciences, Beijing, China, 100190}

\author{Yang Liu} \email{liuyang02@pku.edu.cn}
\affiliation{International Center for Quantum Materials, Peking
	University, Beijing, China, 100871}
\affiliation{Hefei National Laboratory, Hefei, China, 230088}

\author{Xi Lin}
\email{xilin@pku.edu.cn}
\affiliation{International Center for Quantum
	Materials, Peking University, Beijing, China, 100871}
\affiliation{Hefei National Laboratory, Hefei, China, 230088}
\affiliation{Interdisciplinary Institute of Light-Element Quantum Materials and Research Center for Light-Element Advanced Materials, Peking University, Beijing, China, 100871}
	
	\date{\today}
	
	\begin{abstract}
		
	 Surface acoustic wave (SAW) is a powerful technique for investigating quantum phases appearing in two-dimensional electron systems. The electrons respond to the piezoelectric field of SAW through screening, attenuating its amplitude, and shifting its velocity, which is described by the relaxation model. In this work, we systematically study this interaction using orders of magnitude lower SAW amplitude than those in previous studies. At high magnetic fields, when electrons form highly correlated states such as the quantum Hall effect, we observe an anomalously large attenuation of SAW, while the acoustic speed remains considerably high, inconsistent with the conventional relaxation model. This anomaly exists only when the SAW power is sufficiently low.
		
	\end{abstract}                                                                                         
	
	\maketitle
	
	\section{1.Introduction}
	
	Research on charge carrier transport in two-dimensional electron systems (2DES) has unveiled various intriguing quantum phenomena
	\cite{Klitzing.PRL.1980,Tsui.PRL.1982,Jain.CF.2007}.The quantum Hall effect (QHE) is of interest due to its distinctive experimental phenomena: the vanishing of longitudinal resistivity and the precise quantization of Hall resistivity into discrete integers
	\cite{Klitzing.PRL.1980} and fractions
	\cite{Tsui.PRL.1982}. Theoretical explanations for the QHE include the formation of an incompressible quantum liquid that carries superflow current, and disorder-induced localization of sparse quasiparticles\cite{Prange.QHE.1989} At sufficiently large perpendicular magnetic fields, when only several Landau level are occupied, and in samples with sufficiently high mobility, the Coulomb interaction arranges the sparse quasiparticles into an ordered array named the Wigner crystal \cite{Wigner.PR.1934, Lozovik.ZPR.1975,
		Lam.PRB.1984, Levesque.PRB.1984, Willett.PRB.1988,
		Zhu.PRB.1995}. Additionally, the stripe and bubble phases represent two other types of electron solids, emerging as collective charge density waves in the $\ N\ge 1$ spin-resolved Landau levels $\ N\ge 1$ spin-resolved Landau
	levels\cite{Koulakov.PRL.1996, Du.Solid.1999, Lilly.PRL.1999}. The
	interaction between these states and external perturbations, such as
	microwaves\cite{Engel.PRL.1993, Chen.PRL.2003, Lewis.PRB.2005,
		Zhu.PRL.2010, Hatke.NC.2015, Hatke.PRB.2017}, acoustic
	waves\cite{Wixforth.PRL.1986, Wixforth.PRB.1989, Willett.PRL.1990,
		Paalanen.PRB.1992, Willett.PRL.1993, Willett.PRL.2002,
		Friess.NP.2017, Friess.PRL.2018, Friess.PRL.2020, Drichko.APS.2011,
		Drichko.PRB.2015, Drichko.LTP.2017, Wu.arxive.2023},
	light\cite{Arikawa.NP.2017}, in-plane magnetic
	fields\cite{Haug.PRB.1987, Clark.PRL.1989, Engel.PRB.1992,
		Du.PRL.1995, Wang.PRR.2020}, bias voltage\cite{XuebinWang.PRB.2015}, noise\cite{JianSun.FR.2022} and hydrostatic
	pressure\cite{Samkharadze.NP.2016, Schreiber.NC.2018, Huang.PRL.2019}, presents avenues for exploring exotic physics and gaining profound insights into state evolution.
	
	\begin{figure}[!htbp]
		\includegraphics[width=0.48\textwidth]{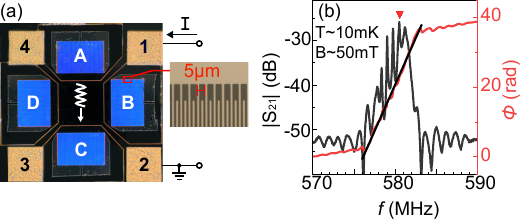}
		\caption{(a) The image of our sample. The
			$d_{\mathrm{m}}\mathrm{=1.2mm}$ square van der Pauw mesa with four
			contacts is defined by wet etching. (b) The frequency spectrum
			measured by IDT pair A \& C. The red triangle marks the resonance
			frequency $f_{c}\approx580.5 $ MHz, at which all the rest
			measurements are performed. }
	\end{figure}
	
	Surface acoustic wave (SAW) provides a contactless approach for
	investigating the transport properties of 2DES. As SAW propagates on
	the surface of GaAs/AlGaAs samples which are piezoelectric, its
	co-propagating piezoelectric field interacts with the adjacent 2DES
	through a relaxation process \cite{Hutson.JAP.1962,
		Bierbaum.APL.1972}. This piezoelectric field will be fully screened
	when the electronic relaxation time (i.e. the transport lift time
	$\tau_{tr}$) is shorter than the acoustic wave period. In the
	weak-coupling limit, the relaxation model predicts that the SAW
	attenuation coefficient $\Gamma$ and its normalized velocity shift
	$\eta = \Delta v/v_{0}$ is related with the 2DES conductivity
	$\sigma_{xx}$ as:
	\begin{eqnarray}
		\Gamma &=& k\frac{K_{\mathrm{eff}}^{2}}{2}\frac{\sigma_{xx}/\sigma_{M}}{1+(\sigma_{xx}/\sigma_{M})^{2}} \\
		\eta &= & \Delta v/v_{0}= \frac{K_{\mathrm{eff}}^{2}}{2}\frac{1}{1+(\sigma_{xx}/\sigma_{M})^{2}}
	\end{eqnarray}
	where
	$\left.v_{0}=v\left(\begin{matrix}{\sigma_{xx}\rightarrow}{\infty}\\\end{matrix}\right.\right)$ and
	${K_{\mathrm{eff}}^{2}}$ is the effective piezoelectric coupling
	constant. The characteristic conductivity
	$\sigma_{M}=v_{0}(\epsilon)\approx 4-7 \times 10^{-7}~\Omega^{-1}$, with $\epsilon$ being the effective dielectric constant.
	
    This relaxation model has been widely used to interpret experimental observations in many previous studies, where the SAW amplitude is large (namely many electrons are confined by the piezo-potential) \cite{Wixforth.PRL.1986, Wixforth.PRB.1989, Willett.PRL.1990,
		Paalanen.PRB.1992, Willett.PRL.1993, Willett.PRL.2002,
		Drichko.APS.2011, Drichko.PRB.2015, Drichko.LTP.2017}.  However, recent experimental observations suggest some insufficiencies when strongly correlated electron solid phases are present	\cite{Friess.NP.2017,Friess.PRL.2018,Friess.PRL.2020,
		Wu.arxive.2023}. In this work, we study 2DES using SAW amplitude that is orders of magnitude smaller than previous studies. Our data
	demonstrate the inadequacy of the relaxation model at low filling
	factors. We find that the interaction between 2DES and SAW is
	influenced by the presence of current and changing acoustic power,
	which are also missed in the relaxation model.
	
	\section{2.Sample and methods}
	
	 Our sample is fabricated from a GaAs/AlGaAs single-interface heterostructure grown by molecular-beam epitaxy whose 2DES layer is located approximately 120\,nm below the surface. The electron density is $1.12\times10^{11}$ cm$^{-2}$ and the low temperature mobility is
	$\mu\sim1\times10^{6}~\mathrm{cm}^{2}\mathrm{V}^{-1}\mathrm{s}^{-1}$.
	The 2DES mesa is a $1.2\times1.2 \, \text{mm}^2$ square at the center,
	with four annealed Ge/Au/Ni/Au Ohmic contacts at the corners of the
	sample. Four interdigital transducers (IDTs), fabricated using
	maskless laser lithography, are symmetrically arranged around the
	square mesa. Each IDT consists of 170 pairs of 1~$\mu$m wide Al finger
	electrodes, separated by an interdigitated gap of 1.5~$\mu$m. Such IDT
	configuration generates SAW with a wavelength of $\lambda=$5~$\mu$m if
	excited at its resonance frequency $f_c=$580.5 MHz; see Fig.l(b). The
	aperture of the IDT spans 1220~$\mu$m, slightly wider than the
	mesa. We removed 200-nm-thick AlGaAs below the IDT, including all doping layers, to ensure optimal performance. All the measurements are
	conducted in a dilution refrigerator whose base temperature is
	approximately $10~\mathrm{mK}$.
	
	The frequency dependence of the amplitude and phase of the
	transmission coefficient, $|\mathbf{S}_{21}|$ and
	$\phi$, is illustrated in Fig.1(b). The data is
	measured at 50~mT and fridge base temperature. The low-field SAW
	velocity can be estimated from the resonance frequency through
	$v_{0}=f_{c}\cdot\lambda\simeq2900~\mathrm{m/s}$, and its total
	traveling time can be deduced from the phase-frequency relation within
	the resonance peak $\tau_0=\partial\phi/\partial(2\pi f)=0.85~\mu $s.
	This estimated SAW travel distance $d=v_0\tau_0\simeq\text{2.5~mm}$ is
	consistent with the 2450~$\mu$m center-to-center distance between
	opposite IDTs.  A homemade RF lock-in amplifier with high sensitivity
	and low noise was utilized to monitor the amplitude and phase shift
	\cite{Wu.arxivLI.2023}. The SAW velocity shift is calculated as
	$\eta =\phi/{2\pi f_{c}\tau}$
	, with $\tau\simeq 0.4~\mu $s, the SAW traveling time through the 1.2
	mm square 2DES mesa \cite{Wu.arxive.2023}. Unless otherwise specified,
	the input excitation power applied to the IDT is $\text{1~nW}$.
	
	\begin{figure}[!htbp]
		\includegraphics[width=0.48\textwidth]{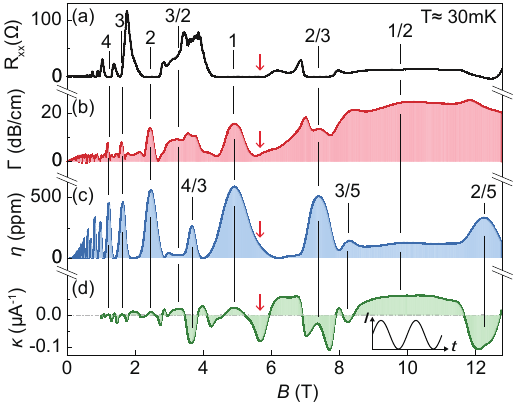}
		\caption{(a-c) The longitudinal resistance $R_{xx}$ measured using
			conventional 4-point transport measurement, the SAW attenuation
			coefficient $\Gamma$ and velocity shift $\eta$ measured when all
			contacts are grounded. (d) The current induced SAW velocity shift
			measured when an oscillating positive current (see the inset)
			flows between contacts 1 and 2.}
	\end{figure}
	
	\section{3.Results and discussion}
	
	Figure 2(a-c) show the longitudinal magneto-resistance $R_{xx}$
	measured using conventional lock-in techniques at $7.3\ \text{Hz}$,
	the SAW attenuation $\Gamma$ and velocity shift $\eta$ in reference
	with their zero magnetic field values. By comparing Fig. 2(a) and (c)
	data, the SAW results are more reliable and sensitive than the
	conventional transport measurement. For example, features
	corresponding to the $\nu=4/3$ fractional quantum Hall effect are
	clearly seen in Fig. 2(c) while they are invisible in Fig. 2(a). Although the response of other phases such as CDW in SAW measurement is not well understood, the observed $\nu=4/3$ feature is more likely a signature of quantum Hall liquid. This is because the formation of CDW phases can either cause anisotropic $\eta$ or a negative $\eta$\cite{Friess.NP.2017}which is not the case here. SAW measurement exhibits features even if droplets of quantum Hall liquid appear, while showing transport features requires the quantum Hall region be sufficiently large to conduct the dissipation-less current. 
	The better performance of SAW measurement can also be seen at low magnetic
	fields, illustrated in Fig. 3(a).  The conductivity of 2DES exhibits
	Shubnikov–de Haas (SdH) oscillations, and the oscillation in $\eta$
	starts at similar field as the best $R_{xx}$. Enhanced $\eta$, seen as
	maximum in Fig. 2(c), appears at even integer filling factors when the
	magnetic field exceeds about 0.1~T. Above 0.38~T, maximum develops at
	odd integer fillings, corresponding to the breaking of spin
	degeneracy. This is also similar to the $R_{xx}$ data. The fact that
	acoustic speed increases at integer fillings is consistent with the
	relaxation model, where the 2DES's screening capability against
	piezoelectric fields vanishes when it forms an incompressible quantum
	Hall liquid. The electrons become mobile when the 2DES is compressible
	at half-integer fillings. They screen the SAW's piezoelectric fields
	and slow down the sound velocity.  It is quite surprising that $\eta$
	almost always equals it's zero-field value, despite the fact that the
	conductivity of 2DES in these high mobility samples vanishes as
	$B^{-2}$\cite{Zhao.CPL.2022}.
	
	\begin{figure}[!htbp] 
		\includegraphics[width=0.48\textwidth]{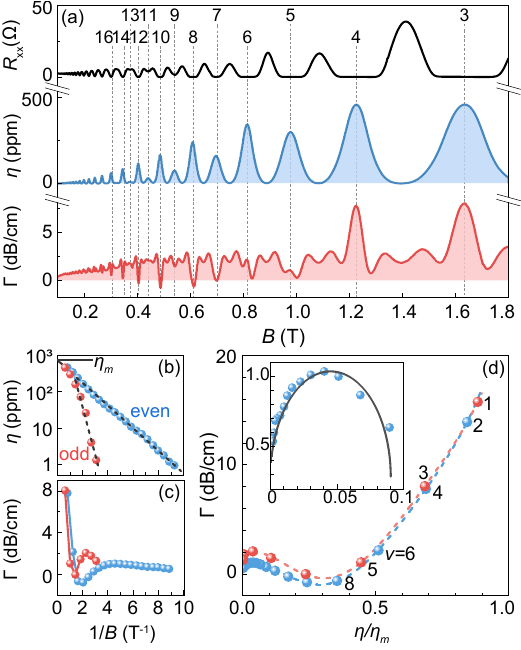}
		\caption{(a) Expanded plot of $\eta$ and $\Gamma$ at low magnetic
			field. (b) $\eta$ vs. $1/B$ exhibits an exponential relation for
			both even and odd integer filling factors. The red/blue dots are
			experiment data at odd/even fillings. The black dashed lines are the
			fitting curves, see text.  (c) $\Gamma$ vs. $1/B$. (d) The relation
			between $\Gamma$ and the normalized velocity shift
			$\eta/\eta_m$. The inset displays the high filling factors
			datapoints and the black line is the best fitting with Eqs.(1) and
			(2) using $K_{\mathrm{eff}}^{2}=1.2\times10^{-4}$.  }
	\end{figure}
	
	We now summarize the $\eta$ and $\Gamma$ at integer fillings as a
	function of their corresponding magnetic fields in Fig. 3(b-c). We
	find that $\eta$ has a clear exponential dependence on $1/\textit{B}$
	over nearly 3 orders of magnitude. Note that the relaxation model is
	not suitable to describe the interaction between the SAW and the
	incompressible QHE when the screening is likely caused by creating
	quasiparticles/quasiholes. One would not be surprised to see the
	exponential relation 	$ \eta \propto \exp(-\Delta_e/\hbar \omega_{c})$  and $\eta \propto \exp(-\Delta_o/E_{Z})$ at odd and even
	filling factors, respectively. By fitting the data in Fig. 3(b), we can
	deduce the energy that describes this SAW-2DES interaction,
	$\Delta_e\simeq 14$K and $\Delta_o\simeq 0.9$K. Here, $\hbar \omega_{c}$ is the cyclotron energy and $E_{Z}$ is the Zeeman energy.
	
	The deviation of our experimental results with the relaxation model is
	more pronounced in
	the evolution of the SAW attenuation $\Gamma$. When we increase the
	magnetic field, $\Gamma$ at integer fillings first increases and reaches
	its maximum at about 0.2 T before decreasing. This is expected by the
	relaxation model and the $\Gamma$ maximum signals that
	$\sigma_{xx}=\sigma_M$. When we further increase the magnetic fields
	to above 0.8~T, a peak in $\Gamma$ appears at integer fillings in
	contrast to the $\Gamma$ minimum seen at low fields. This peak becomes
	dominant at the filling factor $\nu=4$ and continuously grows larger
	as we decrease the filling factor.
	
	The anomalously increasing $\Gamma$  after the low-field peak, seen when $\sigma_{xx}=\sigma_M$ , while $\eta$ monotonically increases, is counterintuitive. The 2DES becomes more incompressible and the
	SAW-2DES interaction should be even weaker for QHE at lower
	fillings. Furthermore, $\Gamma$ seen at low fillings is much larger
	than its low field values in all conditions (below 3 dB/cm in
	Fig. 3(a)). We also note that at $\nu = 1/2,3/2$, where the 2DES is
	compressible and $\eta$ equals its zero field value, $\Gamma$ becomes an
	order of magnitude larger than its low field values; note that the
	unit of $\Gamma$ is logarithmic in Fig. 3. A possible reason could be
	attributed to the emergence of correlated states stabilized by the
	strong electron-electron interactions at low filling factors, such as
	quantum Hall effects, composite fermion Fermi sea, Wigner crystals,
	etc. These states might be effective in damping the SAW's
	piezoelectric fields by mechanisms such as electron-electron
	scattering, while their long range correlation prevents the reducing
	of the acoustic velocity.
	
	We summarize the $\eta$ and $\Gamma$ values for even and odd integer
	filling factors and depicted in Fig. 3(d). The $\Gamma$ vs. the
	normalized velocity shift $\eta/\eta_m$ relation is almost the same
	for the odd and even filling factors. Here,
	$\eta_{\mathrm{m}}\simeq663~\mathrm{ppm}$ is the maximum SAW velocity
	shift at $1/B = 0$ T$^{-1}$ obtained from fitting the Fig. 3(b)
	data. The relaxation model is only applicable to high filling factor
	states, see the Fig. 3(d) inset. At low filling factors, $\Gamma$
	increases nearly linearly with $\eta$, and we do not observe any sign
	of saturation in $\Gamma$ up to $\nu = 1$. As far as we know, a
	suitable theoretical modeling of the SAW-2DES interaction is
	missing. We hope that our clear experimental observations can help to stimulate future investigations.
	
	We would like to emphasize that the above observations can only be
	seen when using extremely low SAW powers, and the presence of a
	conducting current can also vary the observed $\Gamma$ and $\eta$
	\footnote{Recently, Wu et al.\cite{Wu.arxive.2023} revealed an
		interesting discovery: a noticeable modulation in the SAW velocity
		appears when varying current through the 2DES. }. In Fig. 4(c-d), we
	measured the variations in attenuation coefficient
	($\delta\Gamma=\Gamma(I)-\Gamma(0)$) and velocity shift
	($\delta\eta=\eta(I)-\eta(0)$) as a function of DC current
	$I_{\text{DC}}$ at $\nu = 1$ with various input SAW powers. The
	variations of $\Gamma$ and $\eta$ with SAW power when
	$I_{\text{DC}}=0$ are depicted in Fig. 4 (a \& b).
	
	At $\nu=1$, $\Gamma$ decreases by about 15 dB/cm when increasing the
	SAW power from 1 nW to about 80 nW. This is to say, the anomalous
	$\Gamma$ peak seen in Fig. 2(b) can only be seen at sufficiently low
	SAW amplitude. On the other hand, $\Gamma$ slightly increases
	by about 2 dB/cm when we increase the current to almost 1
	$\mu$A. Meanwhile, increasing SAW amplitude reduces $\eta$, while DC
	current leads to an increasing $\eta$. This is rather interesting as
	the data suggests that SAW and current have opposite effect on the
	2DES. One other noteworthy feature in Fig. 4(c) and (d) are the fact
	that the current effect has a clear threshold of about 300 nA in the 1
	nW SAW power trace, which disappears at large SAW power.
	
	In order to compare the DC current effect for different phases, we
	applied a 0.25~Hz, 400 nA peak-to-peak AC current with a 200 nA DC
	offset between contacts 1 and 2 of the sample. The resulting change in
	SAW velocity is represented by a normalization parameter
	$ \kappa=\eta_m^{-1}\cdot(\partial\eta/\partial|I|)$. $\kappa$
	approaches zero when the QHE is strong, e.g. at integer $\nu = 1,2,3,$
	etc. or fractional $\nu = 2/3$. This is expected from the threshold
	feature seen in Fig. 4(b). On the other hand, when the QHE is fragile
	at $\nu = 4/3,2/5$, a substantial negative $\kappa$ is observed,
	suggesting that the applied current makes the 2DES more efficient in
	slowing down the acoustic wave.
	
	In Fig. 2(d), $\kappa$ has a large negative value on the edge of the
	$R_{xx}$ plateaus. In Fig. 4(a-b \& e-f), we focus on one of these negative
	$\kappa$ peaks at 5.7 T (marked by arrows in Fig.2). As we have
	discussed earlier, increasing SAW power and sending current have
	opposite effect on $\eta$ and $\Gamma$ at $\nu=1$, they lead to
	qualitatively the same outcome at 5.7 T.  In both cases, $\eta$
	decreases and $\Gamma$ increases. We note that, the $\Lambda$-shape
	maximum of the $\eta$ vs. $I_{\text{DC}}$ relation at low SAW power
	becomes flat at large SAW power, opposite to the observed trend at
	$\nu=1$. We also would like to point out that the changes in $\eta$ and $\Gamma$
	 tend to saturate at $I_{\text{DC}} \gtrsim 700$ nA, while no
	sign of saturation is seen at $\nu=1$ until 2 $\mu$A, which is still
	far less than the breakdown current of the $\nu=1$ QHE.
	
	\begin{figure}[!htbp]
		\includegraphics[width=0.48\textwidth]{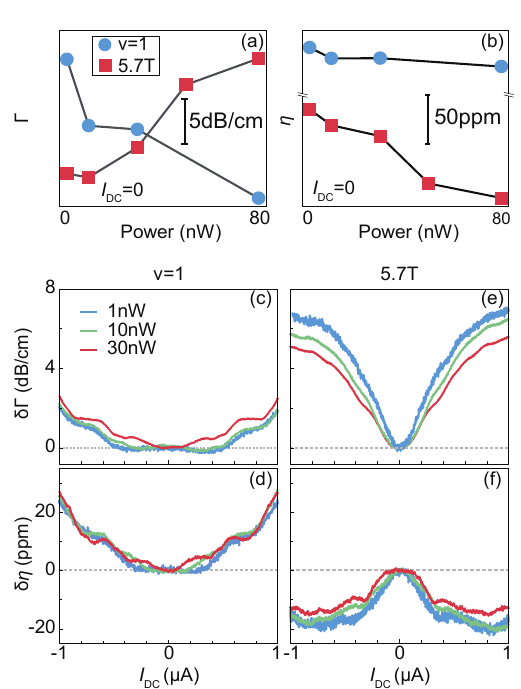}
		\caption{(a-b) The variation of $\Gamma$ and $\eta$ with SAW power at zero
			$I_{\text{DC}}$. (c-f) The current-induced SAW attenuation variation
			$\delta\Gamma$ and velocity shift $\delta\eta$ at $\nu=1$ (left column) and
			$\textit{B}=5.7$~T (right column), using different probing SAW powers.  }
	\end{figure}
	
	\section{4. Conclusion }
	
	Our systematic study shows that the conventional relaxation model is insufficient in describing the SAW-2DES interaction at very low SAW power and when 2DES forms strongly correlated states. We present as much experimental evidence as possible, and hope a comprehensive theoretical models can be proposed in the future.
	
	\section{Acknowledgments }
	\begin{acknowledgments}
		
		We acknowledge support by the National Key Research and Development
		Program of China (Grant No. 2021YFA1401900 and 2019YFA0308403), the Strategic Priority Research Program of the Chinese Academy of Sciences (Grant No. XDB33030000), the National Natural Science Foundation of China (Grant No. 92065104, 12074010 and 12141001) and the Innovation Program for Quantum Science and Technology (Grant No. 2021ZD0302602) for sample fabrication and measurement.
		
	\end{acknowledgments}

	\bibliographystyle{apsrev4-1}
	\bibliography{bib_full_2}

\begin{thebibliography}{48}%
\makeatletter
\providecommand \@ifxundefined [1]{%
 \@ifx{#1\undefined}
}%
\providecommand \@ifnum [1]{%
 \ifnum #1\expandafter \@firstoftwo
 \else \expandafter \@secondoftwo
 \fi
}%
\providecommand \@ifx [1]{%
 \ifx #1\expandafter \@firstoftwo
 \else \expandafter \@secondoftwo
 \fi
}%
\providecommand \natexlab [1]{#1}%
\providecommand \enquote  [1]{``#1''}%
\providecommand \bibnamefont  [1]{#1}%
\providecommand \bibfnamefont [1]{#1}%
\providecommand \citenamefont [1]{#1}%
\providecommand \href@noop [0]{\@secondoftwo}%
\providecommand \href [0]{\begingroup \@sanitize@url \@href}%
\providecommand \@href[1]{\@@startlink{#1}\@@href}%
\providecommand \@@href[1]{\endgroup#1\@@endlink}%
\providecommand \@sanitize@url [0]{\catcode `\\12\catcode `\$12\catcode
  `\&12\catcode `\#12\catcode `\^12\catcode `\_12\catcode `\%12\relax}%
\providecommand \@@startlink[1]{}%
\providecommand \@@endlink[0]{}%
\providecommand \url  [0]{\begingroup\@sanitize@url \@url }%
\providecommand \@url [1]{\endgroup\@href {#1}{\urlprefix }}%
\providecommand \urlprefix  [0]{URL }%
\providecommand \Eprint [0]{\href }%
\providecommand \doibase [0]{http://dx.doi.org/}%
\providecommand \selectlanguage [0]{\@gobble}%
\providecommand \bibinfo  [0]{\@secondoftwo}%
\providecommand \bibfield  [0]{\@secondoftwo}%
\providecommand \translation [1]{[#1]}%
\providecommand \BibitemOpen [0]{}%
\providecommand \bibitemStop [0]{}%
\providecommand \bibitemNoStop [0]{.\EOS\space}%
\providecommand \EOS [0]{\spacefactor3000\relax}%
\providecommand \BibitemShut  [1]{\csname bibitem#1\endcsname}%
\let\auto@bib@innerbib\@empty
\bibitem [{\citenamefont {Klitzing}\ \emph {et~al.}(1980)\citenamefont
  {Klitzing}, \citenamefont {Dorda},\ and\ \citenamefont
  {Pepper}}]{Klitzing.PRL.1980}%
  \BibitemOpen
  \bibfield  {author} {\bibinfo {author} {\bibfnamefont {K.~v.}\ \bibnamefont
  {Klitzing}}, \bibinfo {author} {\bibfnamefont {G.}~\bibnamefont {Dorda}}, \
  and\ \bibinfo {author} {\bibfnamefont {M.}~\bibnamefont {Pepper}},\ }\href
  {\doibase 10.1103/PhysRevLett.45.494} {\bibfield  {journal} {\bibinfo
  {journal} {Phys.Rev.Lett.}\ }\textbf {\bibinfo {volume} {45}},\ \bibinfo
  {pages} {494} (\bibinfo {year} {1980})}\BibitemShut {NoStop}%
\bibitem [{\citenamefont {Tsui}(1982)}]{Tsui.PRL.1982}%
  \BibitemOpen
  \bibfield  {author} {\bibinfo {author} {\bibfnamefont {H.~A.}\ \bibnamefont
  {Tsui}, \bibfnamefont {D.C.and~Stormer}},\ }\href {\doibase
  10.1103/PhysRevLett.48.1559} {\bibfield  {journal} {\bibinfo  {journal}
  {Phys.Rev.Lett.}\ }\textbf {\bibinfo {volume} {48}},\ \bibinfo {pages} {1559}
  (\bibinfo {year} {1982})}\BibitemShut {NoStop}%
\bibitem [{\citenamefont {K.Jain}(2007)}]{Jain.CF.2007}%
  \BibitemOpen
  \bibfield  {author} {\bibinfo {author} {\bibfnamefont {J.}~\bibnamefont
  {K.Jain}},\ }\href {\doibase https://doi.org/10.1017/CBO9780511607561} {\emph
  {\bibinfo {title} {{Composite}{Fermion}}}}\ (\bibinfo  {publisher} {Cambridge
  University Press},\ \bibinfo {year} {2007})\ pp.\ \bibinfo {pages}
  {1--11}\BibitemShut {NoStop}%
\bibitem [{\citenamefont {Prange}\ \emph {et~al.}(1989)\citenamefont {Prange},
  \citenamefont {Girvin} \emph {et~al.}}]{Prange.QHE.1989}%
  \BibitemOpen
  \bibfield  {author} {\bibinfo {author} {\bibfnamefont {R.~E.}\ \bibnamefont
  {Prange}}, \bibinfo {author} {\bibfnamefont {S.~M.}\ \bibnamefont {Girvin}},
  \emph {et~al.},\ }\href {\doibase https://doi.org/10.1007/978-1-4612-3350-3}
  {\emph {\bibinfo {title} {The Quantum Hall Effect [electronic resource]}}}\
  (\bibinfo  {publisher} {New York, NY: Springer US},\ \bibinfo {year} {1989})\
  pp.\ \bibinfo {pages} {1--35}\BibitemShut {NoStop}%
\bibitem [{\citenamefont {Wigner}(1934)}]{Wigner.PR.1934}%
  \BibitemOpen
  \bibfield  {author} {\bibinfo {author} {\bibfnamefont {E.}~\bibnamefont
  {Wigner}},\ }\href {\doibase 10.1103/PhysRev.46.1002} {\bibfield  {journal}
  {\bibinfo  {journal} {Phys. Rev.}\ }\textbf {\bibinfo {volume} {46}},\
  \bibinfo {pages} {1002} (\bibinfo {year} {1934})}\BibitemShut {NoStop}%
\bibitem [{\citenamefont {Lozovik}\ and\ \citenamefont
  {Yudson}(1975)}]{Lozovik.ZPR.1975}%
  \BibitemOpen
  \bibfield  {author} {\bibinfo {author} {\bibfnamefont {Y.~E.}\ \bibnamefont
  {Lozovik}}\ and\ \bibinfo {author} {\bibfnamefont {V.}~\bibnamefont
  {Yudson}},\ }\href@noop {} {\bibfield  {journal} {\bibinfo  {journal} {ZhETF
  Pisma Redaktsiiu}\ }\textbf {\bibinfo {volume} {22}},\ \bibinfo {pages} {26}
  (\bibinfo {year} {1975})}\BibitemShut {NoStop}%
\bibitem [{\citenamefont {Lam}\ and\ \citenamefont
  {Girvin}(1984)}]{Lam.PRB.1984}%
  \BibitemOpen
  \bibfield  {author} {\bibinfo {author} {\bibfnamefont {P.~K.}\ \bibnamefont
  {Lam}}\ and\ \bibinfo {author} {\bibfnamefont {S.~M.}\ \bibnamefont
  {Girvin}},\ }\href {\doibase 10.1103/PhysRevB.30.473} {\bibfield  {journal}
  {\bibinfo  {journal} {Phys. Rev. B}\ }\textbf {\bibinfo {volume} {30}},\
  \bibinfo {pages} {473} (\bibinfo {year} {1984})}\BibitemShut {NoStop}%
\bibitem [{\citenamefont {Levesque}\ \emph {et~al.}(1984)\citenamefont
  {Levesque}, \citenamefont {Weis},\ and\ \citenamefont
  {MacDonald}}]{Levesque.PRB.1984}%
  \BibitemOpen
  \bibfield  {author} {\bibinfo {author} {\bibfnamefont {D.}~\bibnamefont
  {Levesque}}, \bibinfo {author} {\bibfnamefont {J.~J.}\ \bibnamefont {Weis}},
  \ and\ \bibinfo {author} {\bibfnamefont {A.~H.}\ \bibnamefont {MacDonald}},\
  }\href {\doibase 10.1103/PhysRevB.30.1056} {\bibfield  {journal} {\bibinfo
  {journal} {Phys. Rev. B}\ }\textbf {\bibinfo {volume} {30}},\ \bibinfo
  {pages} {1056} (\bibinfo {year} {1984})}\BibitemShut {NoStop}%
\bibitem [{\citenamefont {Willett}\ \emph {et~al.}(1988)\citenamefont
  {Willett}, \citenamefont {Stormer}, \citenamefont {Tsui}, \citenamefont
  {Pfeiffer}, \citenamefont {West},\ and\ \citenamefont
  {Baldwin}}]{Willett.PRB.1988}%
  \BibitemOpen
  \bibfield  {author} {\bibinfo {author} {\bibfnamefont {R.~L.}\ \bibnamefont
  {Willett}}, \bibinfo {author} {\bibfnamefont {H.~L.}\ \bibnamefont
  {Stormer}}, \bibinfo {author} {\bibfnamefont {D.~C.}\ \bibnamefont {Tsui}},
  \bibinfo {author} {\bibfnamefont {L.~N.}\ \bibnamefont {Pfeiffer}}, \bibinfo
  {author} {\bibfnamefont {K.~W.}\ \bibnamefont {West}}, \ and\ \bibinfo
  {author} {\bibfnamefont {K.~W.}\ \bibnamefont {Baldwin}},\ }\href {\doibase
  10.1103/PhysRevB.38.7881} {\bibfield  {journal} {\bibinfo  {journal} {Phys.
  Rev. B}\ }\textbf {\bibinfo {volume} {38}},\ \bibinfo {pages} {7881}
  (\bibinfo {year} {1988})}\BibitemShut {NoStop}%
\bibitem [{\citenamefont {Zhu}\ and\ \citenamefont
  {Louie}(1995)}]{Zhu.PRB.1995}%
  \BibitemOpen
  \bibfield  {author} {\bibinfo {author} {\bibfnamefont {X.}~\bibnamefont
  {Zhu}}\ and\ \bibinfo {author} {\bibfnamefont {S.~G.}\ \bibnamefont
  {Louie}},\ }\href {\doibase 10.1103/PhysRevB.52.5863} {\bibfield  {journal}
  {\bibinfo  {journal} {Phys. Rev. B}\ }\textbf {\bibinfo {volume} {52}},\
  \bibinfo {pages} {5863} (\bibinfo {year} {1995})}\BibitemShut {NoStop}%
\bibitem [{\citenamefont {Koulakov}\ \emph {et~al.}(1996)\citenamefont
  {Koulakov}, \citenamefont {Fogler},\ and\ \citenamefont
  {Shklovskii}}]{Koulakov.PRL.1996}%
  \BibitemOpen
  \bibfield  {author} {\bibinfo {author} {\bibfnamefont {A.~A.}\ \bibnamefont
  {Koulakov}}, \bibinfo {author} {\bibfnamefont {M.~M.}\ \bibnamefont
  {Fogler}}, \ and\ \bibinfo {author} {\bibfnamefont {B.~I.}\ \bibnamefont
  {Shklovskii}},\ }\href {\doibase 10.1103/PhysRevLett.76.499} {\bibfield
  {journal} {\bibinfo  {journal} {Phys. Rev. Lett.}\ }\textbf {\bibinfo
  {volume} {76}},\ \bibinfo {pages} {499} (\bibinfo {year} {1996})}\BibitemShut
  {NoStop}%
\bibitem [{\citenamefont {Du}\ \emph {et~al.}(1999)\citenamefont {Du},
  \citenamefont {Tsui}, \citenamefont {Stormer}, \citenamefont {Pfeiffer},
  \citenamefont {Baldwin},\ and\ \citenamefont {West}}]{Du.Solid.1999}%
  \BibitemOpen
  \bibfield  {author} {\bibinfo {author} {\bibfnamefont {R.}~\bibnamefont
  {Du}}, \bibinfo {author} {\bibfnamefont {D.}~\bibnamefont {Tsui}}, \bibinfo
  {author} {\bibfnamefont {H.}~\bibnamefont {Stormer}}, \bibinfo {author}
  {\bibfnamefont {L.}~\bibnamefont {Pfeiffer}}, \bibinfo {author}
  {\bibfnamefont {K.}~\bibnamefont {Baldwin}}, \ and\ \bibinfo {author}
  {\bibfnamefont {K.}~\bibnamefont {West}},\ }\href
  {http://dx.doi.org/10.1016/s0038-1098(98)00578-x} {\bibfield  {journal}
  {\bibinfo  {journal} {Solid State Communications}\ }\textbf {\bibinfo
  {volume} {109}},\ \bibinfo {pages} {389} (\bibinfo {year}
  {1999})}\BibitemShut {NoStop}%
\bibitem [{\citenamefont {Lilly}\ \emph {et~al.}(1999)\citenamefont {Lilly},
  \citenamefont {Cooper}, \citenamefont {Eisenstein}, \citenamefont
  {Pfeiffer},\ and\ \citenamefont {West}}]{Lilly.PRL.1999}%
  \BibitemOpen
  \bibfield  {author} {\bibinfo {author} {\bibfnamefont {M.~P.}\ \bibnamefont
  {Lilly}}, \bibinfo {author} {\bibfnamefont {K.~B.}\ \bibnamefont {Cooper}},
  \bibinfo {author} {\bibfnamefont {J.~P.}\ \bibnamefont {Eisenstein}},
  \bibinfo {author} {\bibfnamefont {L.~N.}\ \bibnamefont {Pfeiffer}}, \ and\
  \bibinfo {author} {\bibfnamefont {K.~W.}\ \bibnamefont {West}},\ }\href
  {\doibase 10.1103/PhysRevLett.82.394} {\bibfield  {journal} {\bibinfo
  {journal} {Phys. Rev. Lett.}\ }\textbf {\bibinfo {volume} {82}},\ \bibinfo
  {pages} {394} (\bibinfo {year} {1999})}\BibitemShut {NoStop}%
\bibitem [{\citenamefont {Engel}\ \emph {et~al.}(1993)\citenamefont {Engel},
  \citenamefont {Shahar}, \citenamefont {Kurdak},\ and\ \citenamefont
  {Tsui}}]{Engel.PRL.1993}%
  \BibitemOpen
  \bibfield  {author} {\bibinfo {author} {\bibfnamefont {L.~W.}\ \bibnamefont
  {Engel}}, \bibinfo {author} {\bibfnamefont {D.}~\bibnamefont {Shahar}},
  \bibinfo {author} {\bibfnamefont {i.~m.~c.}\ \bibnamefont {Kurdak}}, \ and\
  \bibinfo {author} {\bibfnamefont {D.~C.}\ \bibnamefont {Tsui}},\ }\href
  {\doibase 10.1103/PhysRevLett.71.2638} {\bibfield  {journal} {\bibinfo
  {journal} {Phys. Rev. Lett.}\ }\textbf {\bibinfo {volume} {71}},\ \bibinfo
  {pages} {2638} (\bibinfo {year} {1993})}\BibitemShut {NoStop}%
\bibitem [{\citenamefont {Chen}\ \emph {et~al.}(2003)\citenamefont {Chen},
  \citenamefont {Lewis}, \citenamefont {Engel}, \citenamefont {Tsui},
  \citenamefont {Ye}, \citenamefont {Pfeiffer},\ and\ \citenamefont
  {West}}]{Chen.PRL.2003}%
  \BibitemOpen
  \bibfield  {author} {\bibinfo {author} {\bibfnamefont {Y.}~\bibnamefont
  {Chen}}, \bibinfo {author} {\bibfnamefont {R.~M.}\ \bibnamefont {Lewis}},
  \bibinfo {author} {\bibfnamefont {L.~W.}\ \bibnamefont {Engel}}, \bibinfo
  {author} {\bibfnamefont {D.~C.}\ \bibnamefont {Tsui}}, \bibinfo {author}
  {\bibfnamefont {P.~D.}\ \bibnamefont {Ye}}, \bibinfo {author} {\bibfnamefont
  {L.~N.}\ \bibnamefont {Pfeiffer}}, \ and\ \bibinfo {author} {\bibfnamefont
  {K.~W.}\ \bibnamefont {West}},\ }\href {\doibase
  10.1103/PhysRevLett.91.016801} {\bibfield  {journal} {\bibinfo  {journal}
  {Phys. Rev. Lett.}\ }\textbf {\bibinfo {volume} {91}},\ \bibinfo {pages}
  {016801} (\bibinfo {year} {2003})}\BibitemShut {NoStop}%
\bibitem [{\citenamefont {Lewis}\ \emph {et~al.}(2005)\citenamefont {Lewis},
  \citenamefont {Chen}, \citenamefont {Engel}, \citenamefont {Tsui},
  \citenamefont {Pfeiffer},\ and\ \citenamefont {West}}]{Lewis.PRB.2005}%
  \BibitemOpen
  \bibfield  {author} {\bibinfo {author} {\bibfnamefont {R.~M.}\ \bibnamefont
  {Lewis}}, \bibinfo {author} {\bibfnamefont {Y.~P.}\ \bibnamefont {Chen}},
  \bibinfo {author} {\bibfnamefont {L.~W.}\ \bibnamefont {Engel}}, \bibinfo
  {author} {\bibfnamefont {D.~C.}\ \bibnamefont {Tsui}}, \bibinfo {author}
  {\bibfnamefont {L.~N.}\ \bibnamefont {Pfeiffer}}, \ and\ \bibinfo {author}
  {\bibfnamefont {K.~W.}\ \bibnamefont {West}},\ }\href {\doibase
  10.1103/PhysRevB.71.081301} {\bibfield  {journal} {\bibinfo  {journal} {Phys.
  Rev. B}\ }\textbf {\bibinfo {volume} {71}},\ \bibinfo {pages} {081301}
  (\bibinfo {year} {2005})}\BibitemShut {NoStop}%
\bibitem [{\citenamefont {Zhu}\ \emph {et~al.}(2010)\citenamefont {Zhu},
  \citenamefont {Sambandamurthy}, \citenamefont {Chen}, \citenamefont {Jiang},
  \citenamefont {Engel}, \citenamefont {Tsui}, \citenamefont {Pfeiffer},\ and\
  \citenamefont {West}}]{Zhu.PRL.2010}%
  \BibitemOpen
  \bibfield  {author} {\bibinfo {author} {\bibfnamefont {H.}~\bibnamefont
  {Zhu}}, \bibinfo {author} {\bibfnamefont {G.}~\bibnamefont {Sambandamurthy}},
  \bibinfo {author} {\bibfnamefont {Y.~P.}\ \bibnamefont {Chen}}, \bibinfo
  {author} {\bibfnamefont {P.}~\bibnamefont {Jiang}}, \bibinfo {author}
  {\bibfnamefont {L.~W.}\ \bibnamefont {Engel}}, \bibinfo {author}
  {\bibfnamefont {D.~C.}\ \bibnamefont {Tsui}}, \bibinfo {author}
  {\bibfnamefont {L.~N.}\ \bibnamefont {Pfeiffer}}, \ and\ \bibinfo {author}
  {\bibfnamefont {K.~W.}\ \bibnamefont {West}},\ }\href {\doibase
  10.1103/PhysRevLett.104.226801} {\bibfield  {journal} {\bibinfo  {journal}
  {Phys. Rev. Lett.}\ }\textbf {\bibinfo {volume} {104}},\ \bibinfo {pages}
  {226801} (\bibinfo {year} {2010})}\BibitemShut {NoStop}%
\bibitem [{\citenamefont {Hatke}\ \emph {et~al.}(2015)\citenamefont {Hatke},
  \citenamefont {Liu}, \citenamefont {Engel}, \citenamefont {Shayegan},
  \citenamefont {Pfeiffer}, \citenamefont {West},\ and\ \citenamefont
  {Baldwin}}]{Hatke.NC.2015}%
  \BibitemOpen
  \bibfield  {author} {\bibinfo {author} {\bibfnamefont {A.}~\bibnamefont
  {Hatke}}, \bibinfo {author} {\bibfnamefont {Y.}~\bibnamefont {Liu}}, \bibinfo
  {author} {\bibfnamefont {L.}~\bibnamefont {Engel}}, \bibinfo {author}
  {\bibfnamefont {M.}~\bibnamefont {Shayegan}}, \bibinfo {author}
  {\bibfnamefont {L.}~\bibnamefont {Pfeiffer}}, \bibinfo {author}
  {\bibfnamefont {K.}~\bibnamefont {West}}, \ and\ \bibinfo {author}
  {\bibfnamefont {K.}~\bibnamefont {Baldwin}},\ }\href
  {http://dx.doi.org/10.1038/ncomms8071} {\bibfield  {journal} {\bibinfo
  {journal} {Nature communications}\ }\textbf {\bibinfo {volume} {6}},\
  \bibinfo {pages} {7071} (\bibinfo {year} {2015})}\BibitemShut {NoStop}%
\bibitem [{\citenamefont {Hatke}\ \emph {et~al.}(2017)\citenamefont {Hatke},
  \citenamefont {Liu}, \citenamefont {Engel}, \citenamefont {Pfeiffer},
  \citenamefont {West}, \citenamefont {Baldwin},\ and\ \citenamefont
  {Shayegan}}]{Hatke.PRB.2017}%
  \BibitemOpen
  \bibfield  {author} {\bibinfo {author} {\bibfnamefont {A.~T.}\ \bibnamefont
  {Hatke}}, \bibinfo {author} {\bibfnamefont {Y.}~\bibnamefont {Liu}}, \bibinfo
  {author} {\bibfnamefont {L.~W.}\ \bibnamefont {Engel}}, \bibinfo {author}
  {\bibfnamefont {L.~N.}\ \bibnamefont {Pfeiffer}}, \bibinfo {author}
  {\bibfnamefont {K.~W.}\ \bibnamefont {West}}, \bibinfo {author}
  {\bibfnamefont {K.~W.}\ \bibnamefont {Baldwin}}, \ and\ \bibinfo {author}
  {\bibfnamefont {M.}~\bibnamefont {Shayegan}},\ }\href {\doibase
  10.1103/PhysRevB.95.045417} {\bibfield  {journal} {\bibinfo  {journal} {Phys.
  Rev. B}\ }\textbf {\bibinfo {volume} {95}},\ \bibinfo {pages} {045417}
  (\bibinfo {year} {2017})}\BibitemShut {NoStop}%
\bibitem [{\citenamefont {Wixforth}\ \emph {et~al.}(1986)\citenamefont
  {Wixforth}, \citenamefont {Kotthaus},\ and\ \citenamefont
  {Weimann}}]{Wixforth.PRL.1986}%
  \BibitemOpen
  \bibfield  {author} {\bibinfo {author} {\bibfnamefont {A.}~\bibnamefont
  {Wixforth}}, \bibinfo {author} {\bibfnamefont {J.~P.}\ \bibnamefont
  {Kotthaus}}, \ and\ \bibinfo {author} {\bibfnamefont {G.}~\bibnamefont
  {Weimann}},\ }\href {\doibase 10.1103/PhysRevLett.56.2104} {\bibfield
  {journal} {\bibinfo  {journal} {Phys. Rev. Lett.}\ }\textbf {\bibinfo
  {volume} {56}},\ \bibinfo {pages} {2104} (\bibinfo {year}
  {1986})}\BibitemShut {NoStop}%
\bibitem [{\citenamefont {Wixforth}\ \emph {et~al.}(1989)\citenamefont
  {Wixforth}, \citenamefont {Scriba}, \citenamefont {Wassermeier},
  \citenamefont {Kotthaus}, \citenamefont {Weimann},\ and\ \citenamefont
  {Schlapp}}]{Wixforth.PRB.1989}%
  \BibitemOpen
  \bibfield  {author} {\bibinfo {author} {\bibfnamefont {A.}~\bibnamefont
  {Wixforth}}, \bibinfo {author} {\bibfnamefont {J.}~\bibnamefont {Scriba}},
  \bibinfo {author} {\bibfnamefont {M.}~\bibnamefont {Wassermeier}}, \bibinfo
  {author} {\bibfnamefont {J.~P.}\ \bibnamefont {Kotthaus}}, \bibinfo {author}
  {\bibfnamefont {G.}~\bibnamefont {Weimann}}, \ and\ \bibinfo {author}
  {\bibfnamefont {W.}~\bibnamefont {Schlapp}},\ }\href {\doibase
  10.1103/PhysRevB.40.7874} {\bibfield  {journal} {\bibinfo  {journal} {Phys.
  Rev. B}\ }\textbf {\bibinfo {volume} {40}},\ \bibinfo {pages} {7874}
  (\bibinfo {year} {1989})}\BibitemShut {NoStop}%
\bibitem [{\citenamefont {Willett}\ \emph {et~al.}(1990)\citenamefont
  {Willett}, \citenamefont {Paalanen}, \citenamefont {Ruel}, \citenamefont
  {West}, \citenamefont {Pfeiffer},\ and\ \citenamefont
  {Bishop}}]{Willett.PRL.1990}%
  \BibitemOpen
  \bibfield  {author} {\bibinfo {author} {\bibfnamefont {R.~L.}\ \bibnamefont
  {Willett}}, \bibinfo {author} {\bibfnamefont {M.~A.}\ \bibnamefont
  {Paalanen}}, \bibinfo {author} {\bibfnamefont {R.~R.}\ \bibnamefont {Ruel}},
  \bibinfo {author} {\bibfnamefont {K.~W.}\ \bibnamefont {West}}, \bibinfo
  {author} {\bibfnamefont {L.~N.}\ \bibnamefont {Pfeiffer}}, \ and\ \bibinfo
  {author} {\bibfnamefont {D.~J.}\ \bibnamefont {Bishop}},\ }\href {\doibase
  10.1103/PhysRevLett.65.112} {\bibfield  {journal} {\bibinfo  {journal} {Phys.
  Rev. Lett.}\ }\textbf {\bibinfo {volume} {65}},\ \bibinfo {pages} {112}
  (\bibinfo {year} {1990})}\BibitemShut {NoStop}%
\bibitem [{\citenamefont {Paalanen}\ \emph {et~al.}(1992)\citenamefont
  {Paalanen}, \citenamefont {Willett}, \citenamefont {Littlewood},
  \citenamefont {Ruel}, \citenamefont {West}, \citenamefont {Pfeiffer},\ and\
  \citenamefont {Bishop}}]{Paalanen.PRB.1992}%
  \BibitemOpen
  \bibfield  {author} {\bibinfo {author} {\bibfnamefont {M.~A.}\ \bibnamefont
  {Paalanen}}, \bibinfo {author} {\bibfnamefont {R.~L.}\ \bibnamefont
  {Willett}}, \bibinfo {author} {\bibfnamefont {P.~B.}\ \bibnamefont
  {Littlewood}}, \bibinfo {author} {\bibfnamefont {R.~R.}\ \bibnamefont
  {Ruel}}, \bibinfo {author} {\bibfnamefont {K.~W.}\ \bibnamefont {West}},
  \bibinfo {author} {\bibfnamefont {L.~N.}\ \bibnamefont {Pfeiffer}}, \ and\
  \bibinfo {author} {\bibfnamefont {D.~J.}\ \bibnamefont {Bishop}},\ }\href
  {\doibase 10.1103/PhysRevB.45.11342} {\bibfield  {journal} {\bibinfo
  {journal} {Phys. Rev. B}\ }\textbf {\bibinfo {volume} {45}},\ \bibinfo
  {pages} {11342} (\bibinfo {year} {1992})}\BibitemShut {NoStop}%
\bibitem [{\citenamefont {Willett}\ \emph {et~al.}(1993)\citenamefont
  {Willett}, \citenamefont {Ruel}, \citenamefont {West},\ and\ \citenamefont
  {Pfeiffer}}]{Willett.PRL.1993}%
  \BibitemOpen
  \bibfield  {author} {\bibinfo {author} {\bibfnamefont {R.~L.}\ \bibnamefont
  {Willett}}, \bibinfo {author} {\bibfnamefont {R.~R.}\ \bibnamefont {Ruel}},
  \bibinfo {author} {\bibfnamefont {K.~W.}\ \bibnamefont {West}}, \ and\
  \bibinfo {author} {\bibfnamefont {L.~N.}\ \bibnamefont {Pfeiffer}},\ }\href
  {\doibase 10.1103/PhysRevLett.71.3846} {\bibfield  {journal} {\bibinfo
  {journal} {Phys. Rev. Lett.}\ }\textbf {\bibinfo {volume} {71}},\ \bibinfo
  {pages} {3846} (\bibinfo {year} {1993})}\BibitemShut {NoStop}%
\bibitem [{\citenamefont {Willett}\ \emph {et~al.}(2002)\citenamefont
  {Willett}, \citenamefont {West},\ and\ \citenamefont
  {Pfeiffer}}]{Willett.PRL.2002}%
  \BibitemOpen
  \bibfield  {author} {\bibinfo {author} {\bibfnamefont {R.~L.}\ \bibnamefont
  {Willett}}, \bibinfo {author} {\bibfnamefont {K.~W.}\ \bibnamefont {West}}, \
  and\ \bibinfo {author} {\bibfnamefont {L.~N.}\ \bibnamefont {Pfeiffer}},\
  }\href {\doibase 10.1103/PhysRevLett.88.066801} {\bibfield  {journal}
  {\bibinfo  {journal} {Phys. Rev. Lett.}\ }\textbf {\bibinfo {volume} {88}},\
  \bibinfo {pages} {066801} (\bibinfo {year} {2002})}\BibitemShut {NoStop}%
\bibitem [{\citenamefont {Friess}\ \emph {et~al.}(2017)\citenamefont {Friess},
  \citenamefont {Peng}, \citenamefont {Rosenow}, \citenamefont {von Oppen},
  \citenamefont {Umansky}, \citenamefont {von Klitzing},\ and\ \citenamefont
  {Smet}}]{Friess.NP.2017}%
  \BibitemOpen
  \bibfield  {author} {\bibinfo {author} {\bibfnamefont {B.}~\bibnamefont
  {Friess}}, \bibinfo {author} {\bibfnamefont {Y.}~\bibnamefont {Peng}},
  \bibinfo {author} {\bibfnamefont {B.}~\bibnamefont {Rosenow}}, \bibinfo
  {author} {\bibfnamefont {F.}~\bibnamefont {von Oppen}}, \bibinfo {author}
  {\bibfnamefont {V.}~\bibnamefont {Umansky}}, \bibinfo {author} {\bibfnamefont
  {K.}~\bibnamefont {von Klitzing}}, \ and\ \bibinfo {author} {\bibfnamefont
  {J.~H.}\ \bibnamefont {Smet}},\ }\href {http://dx.doi.org/10.1038/nphys4213}
  {\bibfield  {journal} {\bibinfo  {journal} {Nature Physics}\ }\textbf
  {\bibinfo {volume} {13}},\ \bibinfo {pages} {1124} (\bibinfo {year}
  {2017})}\BibitemShut {NoStop}%
\bibitem [{\citenamefont {Friess}\ \emph {et~al.}(2018)\citenamefont {Friess},
  \citenamefont {Umansky}, \citenamefont {von Klitzing},\ and\ \citenamefont
  {Smet}}]{Friess.PRL.2018}%
  \BibitemOpen
  \bibfield  {author} {\bibinfo {author} {\bibfnamefont {B.}~\bibnamefont
  {Friess}}, \bibinfo {author} {\bibfnamefont {V.}~\bibnamefont {Umansky}},
  \bibinfo {author} {\bibfnamefont {K.}~\bibnamefont {von Klitzing}}, \ and\
  \bibinfo {author} {\bibfnamefont {J.~H.}\ \bibnamefont {Smet}},\ }\href
  {\doibase 10.1103/PhysRevLett.120.137603} {\bibfield  {journal} {\bibinfo
  {journal} {Phys. Rev. Lett.}\ }\textbf {\bibinfo {volume} {120}},\ \bibinfo
  {pages} {137603} (\bibinfo {year} {2018})}\BibitemShut {NoStop}%
\bibitem [{\citenamefont {Friess}\ \emph {et~al.}(2020)\citenamefont {Friess},
  \citenamefont {Dmitriev}, \citenamefont {Umansky}, \citenamefont {Pfeiffer},
  \citenamefont {West}, \citenamefont {von Klitzing},\ and\ \citenamefont
  {Smet}}]{Friess.PRL.2020}%
  \BibitemOpen
  \bibfield  {author} {\bibinfo {author} {\bibfnamefont {B.}~\bibnamefont
  {Friess}}, \bibinfo {author} {\bibfnamefont {I.~A.}\ \bibnamefont
  {Dmitriev}}, \bibinfo {author} {\bibfnamefont {V.}~\bibnamefont {Umansky}},
  \bibinfo {author} {\bibfnamefont {L.}~\bibnamefont {Pfeiffer}}, \bibinfo
  {author} {\bibfnamefont {K.}~\bibnamefont {West}}, \bibinfo {author}
  {\bibfnamefont {K.}~\bibnamefont {von Klitzing}}, \ and\ \bibinfo {author}
  {\bibfnamefont {J.~H.}\ \bibnamefont {Smet}},\ }\href {\doibase
  10.1103/PhysRevLett.124.117601} {\bibfield  {journal} {\bibinfo  {journal}
  {Phys. Rev. Lett.}\ }\textbf {\bibinfo {volume} {124}},\ \bibinfo {pages}
  {117601} (\bibinfo {year} {2020})}\BibitemShut {NoStop}%
\bibitem [{\citenamefont {Drichko}\ \emph {et~al.}(2011)\citenamefont
  {Drichko}, \citenamefont {Smirnov}, \citenamefont {Suslov},\ and\
  \citenamefont {Leadley}}]{Drichko.APS.2011}%
  \BibitemOpen
  \bibfield  {author} {\bibinfo {author} {\bibfnamefont {I.~L.}\ \bibnamefont
  {Drichko}}, \bibinfo {author} {\bibfnamefont {I.~Y.}\ \bibnamefont
  {Smirnov}}, \bibinfo {author} {\bibfnamefont {A.~V.}\ \bibnamefont {Suslov}},
  \ and\ \bibinfo {author} {\bibfnamefont {D.~R.}\ \bibnamefont {Leadley}},\
  }\href {\doibase 10.1103/PhysRevB.83.235318} {\bibfield  {journal} {\bibinfo
  {journal} {Phys. Rev. B}\ }\textbf {\bibinfo {volume} {83}},\ \bibinfo
  {pages} {235318} (\bibinfo {year} {2011})}\BibitemShut {NoStop}%
\bibitem [{\citenamefont {Drichko}\ \emph {et~al.}(2015)\citenamefont
  {Drichko}, \citenamefont {Smirnov}, \citenamefont {Suslov}, \citenamefont
  {Pfeiffer}, \citenamefont {West},\ and\ \citenamefont
  {Galperin}}]{Drichko.PRB.2015}%
  \BibitemOpen
  \bibfield  {author} {\bibinfo {author} {\bibfnamefont {I.~L.}\ \bibnamefont
  {Drichko}}, \bibinfo {author} {\bibfnamefont {I.~Y.}\ \bibnamefont
  {Smirnov}}, \bibinfo {author} {\bibfnamefont {A.~V.}\ \bibnamefont {Suslov}},
  \bibinfo {author} {\bibfnamefont {L.~N.}\ \bibnamefont {Pfeiffer}}, \bibinfo
  {author} {\bibfnamefont {K.~W.}\ \bibnamefont {West}}, \ and\ \bibinfo
  {author} {\bibfnamefont {Y.~M.}\ \bibnamefont {Galperin}},\ }\href {\doibase
  10.1103/PhysRevB.92.205313} {\bibfield  {journal} {\bibinfo  {journal} {Phys.
  Rev. B}\ }\textbf {\bibinfo {volume} {92}},\ \bibinfo {pages} {205313}
  (\bibinfo {year} {2015})}\BibitemShut {NoStop}%
\bibitem [{\citenamefont {Drichko}\ \emph {et~al.}(2017)\citenamefont
  {Drichko}, \citenamefont {Smirnov}, \citenamefont {Suslov}, \citenamefont
  {Galperin}, \citenamefont {Pfeiffer},\ and\ \citenamefont
  {West}}]{Drichko.LTP.2017}%
  \BibitemOpen
  \bibfield  {author} {\bibinfo {author} {\bibfnamefont {I.}~\bibnamefont
  {Drichko}}, \bibinfo {author} {\bibfnamefont {I.~Y.}\ \bibnamefont
  {Smirnov}}, \bibinfo {author} {\bibfnamefont {A.}~\bibnamefont {Suslov}},
  \bibinfo {author} {\bibfnamefont {Y.}~\bibnamefont {Galperin}}, \bibinfo
  {author} {\bibfnamefont {L.}~\bibnamefont {Pfeiffer}}, \ and\ \bibinfo
  {author} {\bibfnamefont {K.}~\bibnamefont {West}},\ }\href
  {https://doi.org/10.1063/1.4975107} {\bibfield  {journal} {\bibinfo
  {journal} {Low Temperature Physics}\ }\textbf {\bibinfo {volume} {43}},\
  \bibinfo {pages} {86} (\bibinfo {year} {2017})}\BibitemShut {NoStop}%
\bibitem [{\citenamefont {Wu}\ \emph {et~al.}(2024{\natexlab{a}})\citenamefont
  {Wu}, \citenamefont {Liu}, \citenamefont {Wang}, \citenamefont {Chung},
  \citenamefont {Gupta}, \citenamefont {Baldwin}, \citenamefont {Pfeiffer},
  \citenamefont {Lin},\ and\ \citenamefont {Liu}}]{Wu.arxive.2023}%
  \BibitemOpen
  \bibfield  {author} {\bibinfo {author} {\bibfnamefont {M.}~\bibnamefont
  {Wu}}, \bibinfo {author} {\bibfnamefont {X.}~\bibnamefont {Liu}}, \bibinfo
  {author} {\bibfnamefont {R.}~\bibnamefont {Wang}}, \bibinfo {author}
  {\bibfnamefont {Y.~J.}\ \bibnamefont {Chung}}, \bibinfo {author}
  {\bibfnamefont {A.}~\bibnamefont {Gupta}}, \bibinfo {author} {\bibfnamefont
  {K.~W.}\ \bibnamefont {Baldwin}}, \bibinfo {author} {\bibfnamefont
  {L.}~\bibnamefont {Pfeiffer}}, \bibinfo {author} {\bibfnamefont
  {X.}~\bibnamefont {Lin}}, \ and\ \bibinfo {author} {\bibfnamefont
  {Y.}~\bibnamefont {Liu}},\ }\href {\doibase 10.1103/PhysRevLett.132.076501}
  {\bibfield  {journal} {\bibinfo  {journal} {Phys. Rev. Lett.}\ }\textbf
  {\bibinfo {volume} {132}},\ \bibinfo {pages} {076501} (\bibinfo {year}
  {2024}{\natexlab{a}})}\BibitemShut {NoStop}%
\bibitem [{\citenamefont {Arikawa}\ \emph {et~al.}(2017)\citenamefont
  {Arikawa}, \citenamefont {Hyodo}, \citenamefont {Kadoya},\ and\ \citenamefont
  {Tanaka}}]{Arikawa.NP.2017}%
  \BibitemOpen
  \bibfield  {author} {\bibinfo {author} {\bibfnamefont {T.}~\bibnamefont
  {Arikawa}}, \bibinfo {author} {\bibfnamefont {K.}~\bibnamefont {Hyodo}},
  \bibinfo {author} {\bibfnamefont {Y.}~\bibnamefont {Kadoya}}, \ and\ \bibinfo
  {author} {\bibfnamefont {K.}~\bibnamefont {Tanaka}},\ }\href
  {https://www.nature.com/articles/nphys4078#citeas} {\bibfield  {journal}
  {\bibinfo  {journal} {Nature Physics}\ }\textbf {\bibinfo {volume} {13}},\
  \bibinfo {pages} {688} (\bibinfo {year} {2017})}\BibitemShut {NoStop}%
\bibitem [{\citenamefont {Haug}\ \emph {et~al.}(1987)\citenamefont {Haug},
  \citenamefont {Klitzing}, \citenamefont {Nicholas}, \citenamefont {Maan},\
  and\ \citenamefont {Weimann}}]{Haug.PRB.1987}%
  \BibitemOpen
  \bibfield  {author} {\bibinfo {author} {\bibfnamefont {R.~J.}\ \bibnamefont
  {Haug}}, \bibinfo {author} {\bibfnamefont {K.~v.}\ \bibnamefont {Klitzing}},
  \bibinfo {author} {\bibfnamefont {R.~J.}\ \bibnamefont {Nicholas}}, \bibinfo
  {author} {\bibfnamefont {J.~C.}\ \bibnamefont {Maan}}, \ and\ \bibinfo
  {author} {\bibfnamefont {G.}~\bibnamefont {Weimann}},\ }\href {\doibase
  10.1103/PhysRevB.36.4528} {\bibfield  {journal} {\bibinfo  {journal} {Phys.
  Rev. B}\ }\textbf {\bibinfo {volume} {36}},\ \bibinfo {pages} {4528}
  (\bibinfo {year} {1987})}\BibitemShut {NoStop}%
\bibitem [{\citenamefont {Clark}\ \emph {et~al.}(1989)\citenamefont {Clark},
  \citenamefont {Haynes}, \citenamefont {Suckling}, \citenamefont {Mallett},
  \citenamefont {Wright}, \citenamefont {Harris},\ and\ \citenamefont
  {Foxon}}]{Clark.PRL.1989}%
  \BibitemOpen
  \bibfield  {author} {\bibinfo {author} {\bibfnamefont {R.~G.}\ \bibnamefont
  {Clark}}, \bibinfo {author} {\bibfnamefont {S.~R.}\ \bibnamefont {Haynes}},
  \bibinfo {author} {\bibfnamefont {A.~M.}\ \bibnamefont {Suckling}}, \bibinfo
  {author} {\bibfnamefont {J.~R.}\ \bibnamefont {Mallett}}, \bibinfo {author}
  {\bibfnamefont {P.~A.}\ \bibnamefont {Wright}}, \bibinfo {author}
  {\bibfnamefont {J.~J.}\ \bibnamefont {Harris}}, \ and\ \bibinfo {author}
  {\bibfnamefont {C.~T.}\ \bibnamefont {Foxon}},\ }\href {\doibase
  10.1103/PhysRevLett.62.1536} {\bibfield  {journal} {\bibinfo  {journal}
  {Phys. Rev. Lett.}\ }\textbf {\bibinfo {volume} {62}},\ \bibinfo {pages}
  {1536} (\bibinfo {year} {1989})}\BibitemShut {NoStop}%
\bibitem [{\citenamefont {Engel}\ \emph {et~al.}(1992)\citenamefont {Engel},
  \citenamefont {Hwang}, \citenamefont {Sajoto}, \citenamefont {Tsui},\ and\
  \citenamefont {Shayegan}}]{Engel.PRB.1992}%
  \BibitemOpen
  \bibfield  {author} {\bibinfo {author} {\bibfnamefont {L.~W.}\ \bibnamefont
  {Engel}}, \bibinfo {author} {\bibfnamefont {S.~W.}\ \bibnamefont {Hwang}},
  \bibinfo {author} {\bibfnamefont {T.}~\bibnamefont {Sajoto}}, \bibinfo
  {author} {\bibfnamefont {D.~C.}\ \bibnamefont {Tsui}}, \ and\ \bibinfo
  {author} {\bibfnamefont {M.}~\bibnamefont {Shayegan}},\ }\href {\doibase
  10.1103/PhysRevB.45.3418} {\bibfield  {journal} {\bibinfo  {journal} {Phys.
  Rev. B}\ }\textbf {\bibinfo {volume} {45}},\ \bibinfo {pages} {3418}
  (\bibinfo {year} {1992})}\BibitemShut {NoStop}%
\bibitem [{\citenamefont {Du}\ \emph {et~al.}(1995)\citenamefont {Du},
  \citenamefont {Yeh}, \citenamefont {Stormer}, \citenamefont {Tsui},
  \citenamefont {Pfeiffer},\ and\ \citenamefont {West}}]{Du.PRL.1995}%
  \BibitemOpen
  \bibfield  {author} {\bibinfo {author} {\bibfnamefont {R.~R.}\ \bibnamefont
  {Du}}, \bibinfo {author} {\bibfnamefont {A.~S.}\ \bibnamefont {Yeh}},
  \bibinfo {author} {\bibfnamefont {H.~L.}\ \bibnamefont {Stormer}}, \bibinfo
  {author} {\bibfnamefont {D.~C.}\ \bibnamefont {Tsui}}, \bibinfo {author}
  {\bibfnamefont {L.~N.}\ \bibnamefont {Pfeiffer}}, \ and\ \bibinfo {author}
  {\bibfnamefont {K.~W.}\ \bibnamefont {West}},\ }\href {\doibase
  10.1103/PhysRevLett.75.3926} {\bibfield  {journal} {\bibinfo  {journal}
  {Phys. Rev. Lett.}\ }\textbf {\bibinfo {volume} {75}},\ \bibinfo {pages}
  {3926} (\bibinfo {year} {1995})}\BibitemShut {NoStop}%
\bibitem [{\citenamefont {Wang}\ \emph {et~al.}(2020)\citenamefont {Wang},
  \citenamefont {Sun}, \citenamefont {Fu}, \citenamefont {Wu}, \citenamefont
  {Chen}, \citenamefont {Pfeiffer}, \citenamefont {West}, \citenamefont {Xie},\
  and\ \citenamefont {Lin}}]{Wang.PRR.2020}%
  \BibitemOpen
  \bibfield  {author} {\bibinfo {author} {\bibfnamefont {P.}~\bibnamefont
  {Wang}}, \bibinfo {author} {\bibfnamefont {J.}~\bibnamefont {Sun}}, \bibinfo
  {author} {\bibfnamefont {H.}~\bibnamefont {Fu}}, \bibinfo {author}
  {\bibfnamefont {Y.}~\bibnamefont {Wu}}, \bibinfo {author} {\bibfnamefont
  {H.}~\bibnamefont {Chen}}, \bibinfo {author} {\bibfnamefont {L.~N.}\
  \bibnamefont {Pfeiffer}}, \bibinfo {author} {\bibfnamefont {K.~W.}\
  \bibnamefont {West}}, \bibinfo {author} {\bibfnamefont {X.~C.}\ \bibnamefont
  {Xie}}, \ and\ \bibinfo {author} {\bibfnamefont {X.}~\bibnamefont {Lin}},\
  }\href {\doibase 10.1103/PhysRevResearch.2.022056} {\bibfield  {journal}
  {\bibinfo  {journal} {Phys. Rev. Res.}\ }\textbf {\bibinfo {volume} {2}},\
  \bibinfo {pages} {022056} (\bibinfo {year} {2020})}\BibitemShut {NoStop}%
\bibitem [{\citenamefont {Wang}\ \emph {et~al.}(2015)\citenamefont {Wang},
  \citenamefont {Fu}, \citenamefont {Du}, \citenamefont {Liu}, \citenamefont
  {Wang}, \citenamefont {Pfeiffer}, \citenamefont {West}, \citenamefont {Du},\
  and\ \citenamefont {Lin}}]{XuebinWang.PRB.2015}%
  \BibitemOpen
  \bibfield  {author} {\bibinfo {author} {\bibfnamefont {X.}~\bibnamefont
  {Wang}}, \bibinfo {author} {\bibfnamefont {H.}~\bibnamefont {Fu}}, \bibinfo
  {author} {\bibfnamefont {L.}~\bibnamefont {Du}}, \bibinfo {author}
  {\bibfnamefont {X.}~\bibnamefont {Liu}}, \bibinfo {author} {\bibfnamefont
  {P.}~\bibnamefont {Wang}}, \bibinfo {author} {\bibfnamefont {L.~N.}\
  \bibnamefont {Pfeiffer}}, \bibinfo {author} {\bibfnamefont {K.~W.}\
  \bibnamefont {West}}, \bibinfo {author} {\bibfnamefont {R.-R.}\ \bibnamefont
  {Du}}, \ and\ \bibinfo {author} {\bibfnamefont {X.}~\bibnamefont {Lin}},\
  }\href {\doibase 10.1103/PhysRevB.91.115301} {\bibfield  {journal} {\bibinfo
  {journal} {Phys. Rev. B}\ }\textbf {\bibinfo {volume} {91}},\ \bibinfo
  {pages} {115301} (\bibinfo {year} {2015})}\BibitemShut {NoStop}%
\bibitem [{\citenamefont {Sun}\ \emph {et~al.}(2022)\citenamefont {Sun},
  \citenamefont {Niu}, \citenamefont {Li}, \citenamefont {Liu}, \citenamefont
  {Pfeiffer}, \citenamefont {West}, \citenamefont {Wang},\ and\ \citenamefont
  {Lin}}]{JianSun.FR.2022}%
  \BibitemOpen
  \bibfield  {author} {\bibinfo {author} {\bibfnamefont {J.}~\bibnamefont
  {Sun}}, \bibinfo {author} {\bibfnamefont {J.}~\bibnamefont {Niu}}, \bibinfo
  {author} {\bibfnamefont {Y.}~\bibnamefont {Li}}, \bibinfo {author}
  {\bibfnamefont {Y.}~\bibnamefont {Liu}}, \bibinfo {author} {\bibfnamefont
  {L.}~\bibnamefont {Pfeiffer}}, \bibinfo {author} {\bibfnamefont
  {K.}~\bibnamefont {West}}, \bibinfo {author} {\bibfnamefont {P.}~\bibnamefont
  {Wang}}, \ and\ \bibinfo {author} {\bibfnamefont {X.}~\bibnamefont {Lin}},\
  }\href {\doibase https://doi.org/10.1016/j.fmre.2021.07.006} {\bibfield
  {journal} {\bibinfo  {journal} {Fundamental Research}\ }\textbf {\bibinfo
  {volume} {2}},\ \bibinfo {pages} {178} (\bibinfo {year} {2022})}\BibitemShut
  {NoStop}%
\bibitem [{\citenamefont {Samkharadze}\ \emph {et~al.}(2016)\citenamefont
  {Samkharadze}, \citenamefont {Schreiber}, \citenamefont {Gardner},
  \citenamefont {Manfra}, \citenamefont {Fradkin},\ and\ \citenamefont
  {Csáthy}}]{Samkharadze.NP.2016}%
  \BibitemOpen
  \bibfield  {author} {\bibinfo {author} {\bibfnamefont {N.}~\bibnamefont
  {Samkharadze}}, \bibinfo {author} {\bibfnamefont {K.~A.}\ \bibnamefont
  {Schreiber}}, \bibinfo {author} {\bibfnamefont {G.~C.}\ \bibnamefont
  {Gardner}}, \bibinfo {author} {\bibfnamefont {M.~J.}\ \bibnamefont {Manfra}},
  \bibinfo {author} {\bibfnamefont {E.}~\bibnamefont {Fradkin}}, \ and\
  \bibinfo {author} {\bibfnamefont {G.~A.}\ \bibnamefont {Csáthy}},\ }\href
  {\doibase 10.1038/nphys3523} {\bibfield  {journal} {\bibinfo  {journal}
  {Nature Physics}\ }\textbf {\bibinfo {volume} {12}},\ \bibinfo {pages} {191}
  (\bibinfo {year} {2016})}\BibitemShut {NoStop}%
\bibitem [{\citenamefont {Schreiber}\ \emph {et~al.}(2018)\citenamefont
  {Schreiber}, \citenamefont {Samkharadze}, \citenamefont {Gardner},
  \citenamefont {Lyanda-Geller}, \citenamefont {Manfra}, \citenamefont
  {Pfeiffer}, \citenamefont {West},\ and\ \citenamefont
  {Csáthy}}]{Schreiber.NC.2018}%
  \BibitemOpen
  \bibfield  {author} {\bibinfo {author} {\bibfnamefont {K.~A.}\ \bibnamefont
  {Schreiber}}, \bibinfo {author} {\bibfnamefont {N.}~\bibnamefont
  {Samkharadze}}, \bibinfo {author} {\bibfnamefont {G.~C.}\ \bibnamefont
  {Gardner}}, \bibinfo {author} {\bibfnamefont {Y.}~\bibnamefont
  {Lyanda-Geller}}, \bibinfo {author} {\bibfnamefont {M.~J.}\ \bibnamefont
  {Manfra}}, \bibinfo {author} {\bibfnamefont {L.~N.}\ \bibnamefont
  {Pfeiffer}}, \bibinfo {author} {\bibfnamefont {K.~W.}\ \bibnamefont {West}},
  \ and\ \bibinfo {author} {\bibfnamefont {G.~A.}\ \bibnamefont {Csáthy}},\
  }\href {\doibase 10.1038/s41467-018-04879-1} {\bibfield  {journal} {\bibinfo
  {journal} {Nature Communications}\ }\textbf {\bibinfo {volume} {9}},\
  \bibinfo {pages} {2400} (\bibinfo {year} {2018})}\BibitemShut {NoStop}%
\bibitem [{\citenamefont {Huang}\ \emph {et~al.}(2019)\citenamefont {Huang},
  \citenamefont {Wang}, \citenamefont {Pfeiffer}, \citenamefont {West},
  \citenamefont {Baldwin}, \citenamefont {Liu},\ and\ \citenamefont
  {Lin}}]{Huang.PRL.2019}%
  \BibitemOpen
  \bibfield  {author} {\bibinfo {author} {\bibfnamefont {K.}~\bibnamefont
  {Huang}}, \bibinfo {author} {\bibfnamefont {P.}~\bibnamefont {Wang}},
  \bibinfo {author} {\bibfnamefont {L.~N.}\ \bibnamefont {Pfeiffer}}, \bibinfo
  {author} {\bibfnamefont {K.~W.}\ \bibnamefont {West}}, \bibinfo {author}
  {\bibfnamefont {K.~W.}\ \bibnamefont {Baldwin}}, \bibinfo {author}
  {\bibfnamefont {Y.}~\bibnamefont {Liu}}, \ and\ \bibinfo {author}
  {\bibfnamefont {X.}~\bibnamefont {Lin}},\ }\href {\doibase
  10.1103/PhysRevLett.123.206602} {\bibfield  {journal} {\bibinfo  {journal}
  {Phys. Rev. Lett.}\ }\textbf {\bibinfo {volume} {123}},\ \bibinfo {pages}
  {206602} (\bibinfo {year} {2019})}\BibitemShut {NoStop}%
\bibitem [{\citenamefont {Hutson}\ and\ \citenamefont
  {White}(1962)}]{Hutson.JAP.1962}%
  \BibitemOpen
  \bibfield  {author} {\bibinfo {author} {\bibfnamefont {A.}~\bibnamefont
  {Hutson}}\ and\ \bibinfo {author} {\bibfnamefont {D.~L.}\ \bibnamefont
  {White}},\ }\href {https://doi.org/10.1063/1.1728525} {\bibfield  {journal}
  {\bibinfo  {journal} {Journal of Applied Physics}\ }\textbf {\bibinfo
  {volume} {33}},\ \bibinfo {pages} {40} (\bibinfo {year} {1962})}\BibitemShut
  {NoStop}%
\bibitem [{\citenamefont {Bierbaum}(1972)}]{Bierbaum.APL.1972}%
  \BibitemOpen
  \bibfield  {author} {\bibinfo {author} {\bibfnamefont {P.}~\bibnamefont
  {Bierbaum}},\ }\href {https://doi.org/10.1063/1.1654269} {\bibfield
  {journal} {\bibinfo  {journal} {Applied Physics Letters}\ }\textbf {\bibinfo
  {volume} {21}},\ \bibinfo {pages} {595} (\bibinfo {year} {1972})}\BibitemShut
  {NoStop}%
\bibitem [{\citenamefont {Wu}\ \emph {et~al.}(2024{\natexlab{b}})\citenamefont
  {Wu}, \citenamefont {Liu}, \citenamefont {Wang}, \citenamefont {Lin},\ and\
  \citenamefont {Liu}}]{Wu.arxivLI.2023}%
  \BibitemOpen
  \bibfield  {author} {\bibinfo {author} {\bibfnamefont {M.}~\bibnamefont
  {Wu}}, \bibinfo {author} {\bibfnamefont {X.}~\bibnamefont {Liu}}, \bibinfo
  {author} {\bibfnamefont {R.}~\bibnamefont {Wang}}, \bibinfo {author}
  {\bibfnamefont {X.}~\bibnamefont {Lin}}, \ and\ \bibinfo {author}
  {\bibfnamefont {Y.}~\bibnamefont {Liu}},\ }\href
  {http://dx.doi.org/10.48550/arXiv.2311.01718} {\bibfield  {journal} {\bibinfo
   {journal} {arXiv:2311.01718 [physics.app-ph]}\ } (\bibinfo {year}
  {2024}{\natexlab{b}})}\BibitemShut {NoStop}%
\bibitem [{\citenamefont {Zhao}\ \emph {et~al.}(2022)\citenamefont {Zhao},
  \citenamefont {Lin}, \citenamefont {Chung}, \citenamefont {Baldwin},
  \citenamefont {Pfeiffer},\ and\ \citenamefont {Liu}}]{Zhao.CPL.2022}%
  \BibitemOpen
  \bibfield  {author} {\bibinfo {author} {\bibfnamefont {L.}~\bibnamefont
  {Zhao}}, \bibinfo {author} {\bibfnamefont {W.}~\bibnamefont {Lin}}, \bibinfo
  {author} {\bibfnamefont {Y.~J.}\ \bibnamefont {Chung}}, \bibinfo {author}
  {\bibfnamefont {K.~W.}\ \bibnamefont {Baldwin}}, \bibinfo {author}
  {\bibfnamefont {L.~N.}\ \bibnamefont {Pfeiffer}}, \ and\ \bibinfo {author}
  {\bibfnamefont {Y.}~\bibnamefont {Liu}},\ }\href {\doibase
  10.1088/0256-307X/39/9/097301} {\bibfield  {journal} {\bibinfo  {journal}
  {Chinese Physics Letters}\ }\textbf {\bibinfo {volume} {39}},\ \bibinfo {eid}
  {97301} (\bibinfo {year} {2022})}\BibitemShut {NoStop}%
\bibitem [{Note1()}]{Note1}%
  \BibitemOpen
  \bibinfo {note} {Recently, Wu et al.\cite {Wu.arxive.2023} revealed an
  interesting discovery: a noticeable modulation in the SAW velocity appears
  when varying current through the 2DES.}\BibitemShut {Stop}%
\end{thebibliography}%

\end{document}